\documentclass[twocolumn]{aastex631}

\usepackage{soul}

\begin{document}
 
\title{IPHAS J190812.63+045728.1: A deeply eclipsing, X-ray bright cataclysmic variable with frequent outbursts}

\author[0000-0002-6211-7226]{Raimundo Lopes de Oliveira}
\affiliation{Departamento de F\'isica, Universidade Federal de Sergipe, 
Av. Marechal Rondon, S/N, 49100-000, S\~ao Crist\'ov\~ao, SE, Brazil}
\affiliation{Observat\'orio Nacional, Rua Gal. Jos\'e Cristino 77, 20921-400, 
Rio~de~Janeiro, RJ, Brazil}

\author{Albert Bruch}
\affiliation{Laborat\'orio Nacional de Astrof\'isica, Rua Estados Unidos, 154, 
CEP 37504-364, Itajub\'a, MG, Brazil}

\author[0000-0002-8286-8094]{Koji Mukai}
\affiliation{CRESST II and X-ray Astrophysics Laboratory, NASA/GSFC, 
Greenbelt, MD 20771, USA}
\affiliation{Department of Physics, University of Maryland, Baltimore County, 
1000 Hilltop Circle, Baltimore, MD 21250, USA}

\author[0009-0001-7276-3571]{Amanda S. de Araújo}
\affiliation{Departamento de F\'isica, Universidade Federal de Sergipe, 
Av. Marechal Rondon, S/N, 49100-000, S\~ao Crist\'ov\~ao, SE, Brazil}

\author{Guilherme Parisotto Guimarães}
\affiliation{Departamento de F\'isica, Universidade Federal de Sergipe, 
Av. Marechal Rondon, S/N, 49100-000, S\~ao Crist\'ov\~ao, SE, Brazil}

\author[0000-0001-8179-1147]{Ted Leandro de Almeida}
\affiliation{Instituto Nacional de Pesquisas Espaciais, Avenida dos Astronautas, 1.758, Jardim Granja, S\~{a}o Jos\'{e} dos Campos, S\~{a}o Paulo, Brazil}
\affiliation{Laborat\'orio Nacional de Astrof\'isica, Rua Estados Unidos, 154,
CEP 37504-364, Itajub\'a, MG, Brazil}

\begin{abstract}
We present a multi-wavelength characterization of IPHAS~J190812.63+045728.1. 
Independently discovered by us after being serendipitously observed by 
XMM-{\it Newton}, followed up with optical photometry with SPARC4/OPD and 
spectroscopy with GMOS/Gemini South, and exploring archival TESS and ZTF observations, we reveal the system to be an eclipsing 
accreting white dwarf (WD) system with an orbital period of 5.073\,h. The deep optical and flat-bottomed 
X-ray eclipses completely occult the WD and the accretion disk, 
providing a model-independent, uncontaminated view of the donor star which we 
classify as an M2.3 dwarf with a mass of $\sim 0.50\,M_\odot$. The system is a modestly 
luminous ($L_{\rm X\,;\,0.3-10\,keV} \sim 1.7 \times 10^{32}$\,erg\,s$^{-1}$)
hard thermal (consistent with a shock-heated plasma cooling down from $kT \sim 26$\,keV) X-ray source. While these X-ray properties are typical
of low luminosity intermediate polars (IPs), long-term 
archival photometry reveals frequent, large-amplitude outbursts 
($\Delta m \sim 2$\,mag in the ZTF $zr$ band) that are the typical characteristics of dwarf novae, and would be unusual for IPs. 
The weakness of the He\,\textsc{II}\,$\lambda$4686 emission line does not allow us to decide between the IP and the dwarf nova scenarios. On the other hand, the optical and X-ray light curves do not exhibit periodic variations caused by the WD rotation. 
If the system is indeed an IP, the accretion geometry is likely unusual while the dwarf nova-like outbursts indicating disk instabilities would make IPHAS~J190812.63+045728.1 a benchmark for testing disk instability models in the presence of magnetically truncated accretion disks. Otherwise, the system is a dwarf nova with X-ray properties at the extreme end of its class.

\end{abstract}


\section{Introduction} \label{sec:intro}

Cataclysmic variables (CVs) are close interacting binary systems consisting 
of a white dwarf (WD) primary and a late-type main-sequence secondary star. 
The evolution and emission properties of these systems are primarily driven 
by mass transfer from the secondary via Roche-lobe overflow, making them 
excellent astrophysical laboratories for studying accretion physics 
\citep[e.g.,][]{Mukai_2017}. In recent years, large-scale time-domain and 
spectroscopic surveys have significantly expanded the known population of CVs, 
uncovering systems with diverse photometric and spectral behaviors.

One such system is IPHAS~J190812.63+045728.1 (hereafter referred to as
IPHAS~J1908; also known as ZTF18abciqza). The source was initially 
identified as a promising CV candidate by the Zwicky Transient Facility 
(ZTF) owing to its significant time variability \citep[$\Delta$mag = 2.8;][]{Szkody20}. 
The astrometric and photometric properties of the system have 
been further constrained by the \textit{Gaia} mission. The source is listed 
in Gaia EDR3 \citep[EDR3 ID 4305450147247974784;][]{GaiaColl16} with mean 
magnitudes of $G = 18.05 \pm 0.04$, $BP = 18.60 \pm 0.14$, and 
$RP = 17.07 \pm 0.10$. 
Furthermore, based on Gaia parallaxes, 
\citet{Bailer-Jones21} estimated its geometric distance to be 
$1076^{+181}_{-124}$ pc, securely identifying it as a Galactic source.

Despite the identification from optical photometry, fully understanding the 
accretion mechanism and assigning IPHAS~J1908 to a specific CV class 
requires multi-wavelength analysis. Optical and X-ray 
spectroscopy are valuable to this end. The X-ray emission in these objects 
typically originates from shock-heated gas in an accretion column or 
a boundary layer, and is also the origin to optical features through 
reprocessing in irradiated 
material. In this work, we present an analysis of the system focusing on 
its optical and X-ray photometric and spectroscopic features. We explore 
optical spectroscopy from the Gemini Multi-Object Spectrographs (GMOS) on Gemini South and optical photometry from the  Simultaneous Polarimeter and Rapid Camera in 4 bands \citep[SPARC4;][]{2012SPIE.8446E..26R,2025PASP..137c5003B}
 mounted in the 1.6\,m telescope at the Pico dos Dias Observatory (OPD) in Brazil. We also investigate archival X-ray photometry and spectroscopy from 
the XMM-{\it Newton}, the start point of this work, alongside long-term optical photometry 
from the Transiting Exoplanet Survey Satellite \citep[TESS;][]{2015JATIS...1a4003R} and Zwicky Transient Facility \citep[ZTF;][] {Bellm_2019,Masci_2019} databases.

\section{Observations and Methodology}
\label{Observations and Methodology}

\subsection{Zwicky Transient Facility (ZTF)}
\label{ZTF}

\begin{figure}[ht!]
\figurenum{1}
\centering
\includegraphics[trim={1.5cm 3cm 1.5cm 7.8cm}, clip, width=\columnwidth]{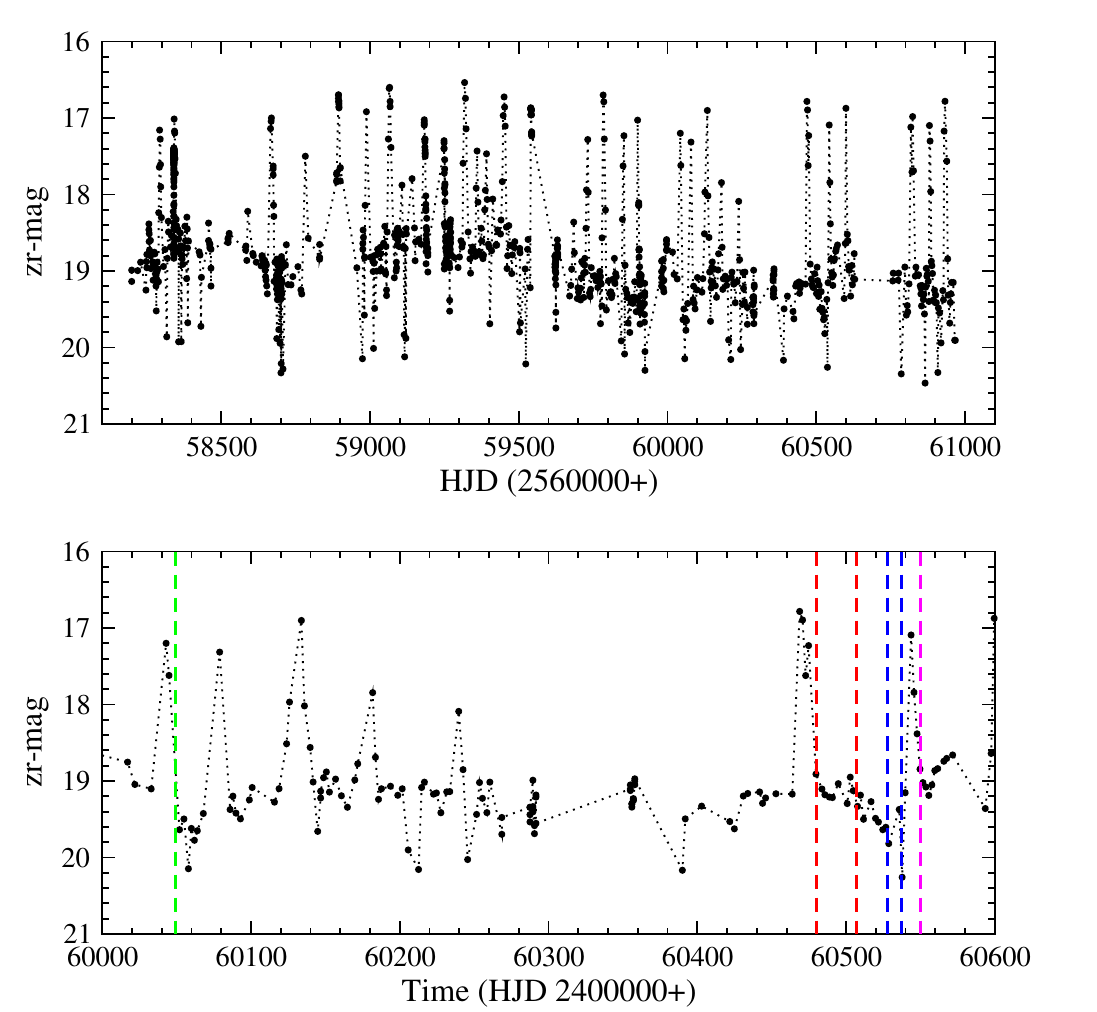}
\caption{{\it Top:} The ZTF light curve of IPHAS~J1908 in the $zr$ band.
{\it Bottom:} An expanded part of the light curve where the intervals 
of the TESS (red) and OPD (blue) photometry, as well as the 
epochs of the spectroscopic (magenta) and X-ray (green) observations
are marked with broken vertical lines.}
\label{ztf-lc}
\end{figure}

Within the time interval from
November 21, 2018 to October 19, 2025, IPHAS~J1908 was observed 905 times
with a mean cadence of 3.10 days in the $zr$ band by ZTF \citep{Bellm_2019,Masci_2019}. The upper panel of Fig.~\ref{ztf-lc} shows the entire light curve. It
is characterized by frequent outbursts with an amplitude of $\sim$\,2 mag 
superposed upon a slightly varying quiescent level. The number of data points 
below this level is compatible with randomly distributed observations 
coinciding with the narrow eclipses observed in IPHAS~J1908 
(see Section~\ref{Timing investigation}). The light curve is sufficiently well sampled to permit an estimate of the average outburst recurrence period and the duty cycle, defined as ratio between the duration of an outburst and the time difference between its start and the beginning of the next outburst. We find a mean recurrence period of 60(22) d (median: 55 d), ranging between 32 and 114 d. The average duty cycle is 40(15) \% (median: 39 \%) and ranges between 22 \% and 75 \%. In both cases, the distribution is skewed to small values.  
The lower panel of Fig.~\ref{ztf-lc} 
presents an 
expanded part of the light curve where we mark with dashed vertical
lines the intervals of the TESS (red) and OPD (blue) 
photometry, as well as the epochs of the spectroscopic (magenta) and 
 X-ray (green) observations.

\subsection{XMM-{\it Newton}}
\label{sct:xmm}

IPHAS J1908 was serendipitously observed as a faint, off-axis source ($\sim$\,10.5 arcmin) by the XMM-\textit{Newton} EPIC cameras on April 15, 
2023, during an exposure lasting $\sim$\,102\,ks  (ObsID: 0904520101; PI: K. Kayama; Observation Mode: PrimeFullWindow; Filter: medium). 
We used the EPIC-pn data 
to identify periods of high solar particle background. Because these background flares occurred only toward the end of the observation, we obtained a clean good time interval of $\sim$ 93.1\,ks. The event list for all three EPIC camera were then filtered using the same time intervals. The X-ray (J2000) coordinates of IPHAS J1908 are RA\,=\,19$^{\rm h}$08$^{\rm m}$12.65$^{\rm s}$ 
and Dec.\,=\,+04$^\circ$57$^\prime$28.9$^{\prime\prime}$, with an angular 
distance of 0.88 arcsec from the \textit{Gaia} EDR3 optical counterpart. As is indicated by the green 
vertical line in Fig.~\ref{ztf-lc}, the system was on the decline from or just after an outburst during these observations.

The XMM-{\it Newton} data were reduced following standard procedures with 
the Science Analysis System (SAS) v22.1.0, using Current Calibration 
Files available as of April 5, 2026. 
Data were reprocessed with the \textsc{epproc} and \textsc{emproc} SAS tasks 
for the EPIC-pn and MOS1/MOS2 cameras, respectively. 

Source events for spectra and light curves were extracted from a circular region with a radius of 12.5 arcsec centered on 
the target, while background events were extracted from a nearby source-free 
circular region of equal radius on the same CCD. The mean net count rates 
(0.6-7.8 keV) are estimated to be 
$0.0368 \pm 0.0008$ count\,s$^{-1}$ for pn, 
$0.0128 \pm 0.0004$ count\,s$^{-1}$ for MOS1, and 
$0.0123 \pm 0.0004$ count\,s$^{-1}$ for MOS2, with the source accounting for approximately 
88\% of the total counts in the extraction regions across the three 
cameras.  

\subsection{Gemini South}
\label{Gemini South}

Long-slit optical spectroscopy of IPHAS~J1908 was obtained using the 
Gemini Multi-Object Spectrograph (GMOS) mounted on the Gemini South 
telescope, under the Director's Discretionary Time (DDT) program 
GS-2024B-DD-103 (PI: R. Lopes de Oliveira). The target was observed on 
August 27, 2024 during the end of the decline from an outburst 
(see the magenta line in Fig.~\ref{ztf-lc}), 
comprising four 1200\,s exposures with the B480 grating 
centered at 5100\,\AA, and another four 1200\,s exposures centered at 
5200\,\AA\ to bridge the CCD gaps and bad columns. The first observation 
was discarded due to technical issues. For the remaining exposures, the 
airmass varied from 1.61 to 1.28. A slit width of 1.0 arcsec was used. 

Data reduction was performed using the  Data Reduction 
for Astronomy from Gemini Observatory North and South software package 
\citep[\textsc{dragons} v3.1.0;][]{2023PASP..135g4502L}, following standard procedures, including bias 
subtraction, flat-fielding, wavelength calibration, and flux calibration 
using the photometric standard star LTT 7379 (a G0 star with 
V\,$\sim$\,10.2~mag), which was observed for 60\,s on August 25, 2024, at 
an airmass of 1.10. 

\begin{figure*}[ht!]
\figurenum{2}
\centering
\includegraphics[trim={0cm 0cm 0cm 0cm}, clip, width=\textwidth]{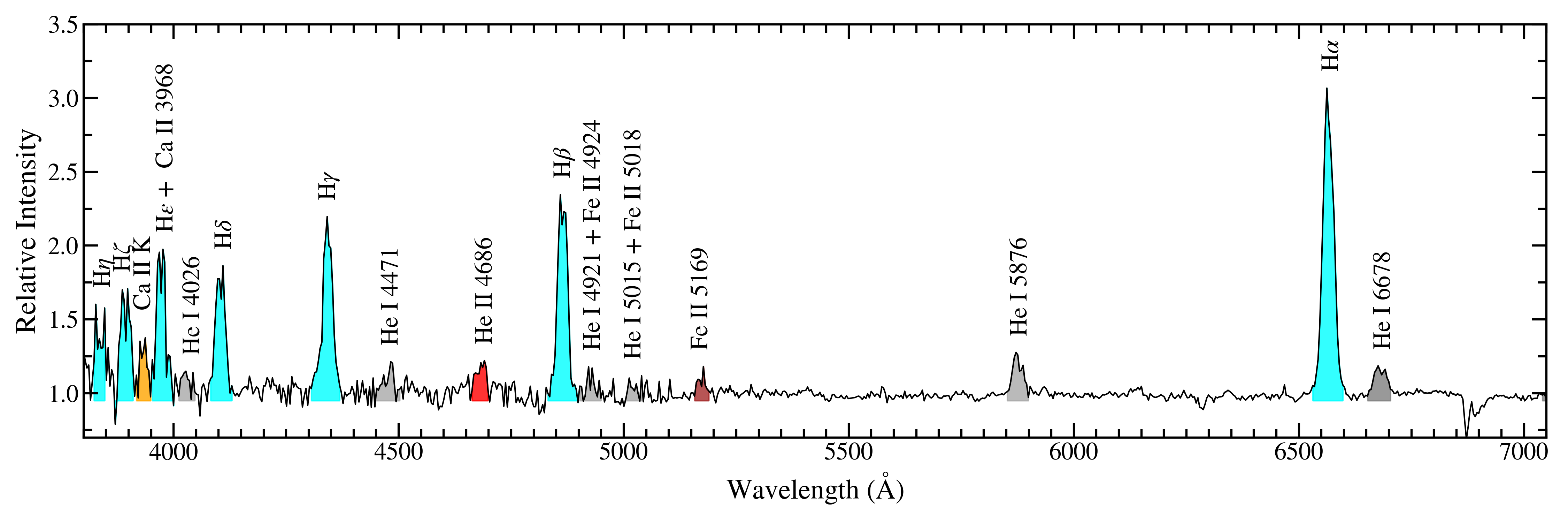}
   \caption{Optical spectrum of IPHAS~J1908 taken with GMOS at Gemini 
   South. Because of calibration uncertainties at the extrema, the spectrum
   was normalized to the continuum.}
    \label{fig:opticalspectrum}
\end{figure*}

The combined spectrum of IPHAS J1908 covers the wavelength range from 
3585\,\AA\ to 7182\,\AA. However, due to poor flux calibration at the 
extreme red and blue ends, we restricted our analysis to the 
3800--7050\,\AA\ region, which contains the relevant spectral features 
for this work (Fig. \ref{fig:opticalspectrum}).

\subsection{Pico dos Dias Observatory}
\label{sct:opd}

\begin{table}[t]
\centering
\begin{tabular}{ccccc}
\hline
Date & \multicolumn{2}{c}{Time (UT)} & Time Resolution & Telescope \\
\cline{2-3}
 (in 2024)    & Start       & End       & (sec)           &        \\
\hline
Aug.\ 05/06 & 23.73           & 5.09 &           183 & ZE \\
Aug.\ 06    & \phantom{0}0.63 & 4.72 & \phantom{0}61 & PE \\
Aug.\ 06/07 &           22.69 & 5.55 &           183 & ZE \\
Aug.\ 07/08 &           22.05 & 3.92 &           183 & ZE \\
Aug.\ 11/12 &           22.92 & 4.54 & \phantom{0}61 & PE \\
Aug.\ 12/13 &           22.03 & 4.73 & \phantom{0}81 & PE \\
Aug.\ 13/14 &           22.14 & 4.71 & \phantom{0}81 & PE \\
\hline
\end{tabular}
    \caption{OPD: journal of the observations}
    \label{tab:opdobs}
\end{table}

Optical photometry of IPHAS~J1908 was obtained using the 1.6\,m Perkin-Elmer 
(PE) and 0.6\,m Zeiss (ZE) telescopes at OPD between August 5 and 14, 2024 
(as indicated by the blue lines in Fig.~\ref{ztf-lc}).
Observations with the ZE telescope were carried out in white light 
(unfiltered), whereas the SPARC4 camera mounted on the PE telescope 
enabled simultaneous imaging in four photometric bands. 

While sky conditions were largely photometric throughout the observing run, 
specific intervals affected by passing clouds were removed from the final 
light curves. Differential aperture photometry was performed using the star 
NOMAD1~0949-0448628 as the reference for the ZE data, and an average of an 
ensemble of field comparison stars for the PE data. Their average 
magnitude permits to transform the differential magnitude scale to
a calibrated scale. A log of 
the photometric observations is provided in Table \ref{tab:opdobs}. 
IPHAS-J1908 remained in 
quiescence during the entire observing mission, at an average out-of-eclipse $r$ band magnitude of $\sim$19.6~mag which is in agreement with the ZTF light curve (Fig.~\ref{ztf-lc}). After correction
for the substantial interstellar extinction towards the system (see Section~\ref{Discussion}), we obtained intrinsic colors of $g-r$ = 0.78, $r-i$ = 0.51, and $i-z$ = 0.21. 

\subsection{Transiting Exoplanet Survey Satellite
 (TESS)}
\label{TESS}

The TESS mission \citep{Ricker14} performed photometric observations 
of IPHAS~J1908 (named as TIC~1790033760 in TESS) in Sector 80 during 26.45 days, from June 18 through July 15, 2024, with a cadence of 2 minutes. These archival data were downloaded from
the Barbara A.\ Mikulski Archive for Space Telescopes.\footnote{MAST, 
https://archive.stsci.edu}

The analysis was performed using Pre-Search Data Conditioning Simple 
Aperture Photometry (PDCSAP) data, which are corrected for instrumental 
systematics and long-term trends inherent to TESS photometry. This 
correction significantly improves the signal-to-noise ratio and removes 
variability on longer time scales ($>$1~d) not associated with the 
intrinsic behavior of the target but leaves short time scale ($<$1~d) 
variations untouched \citep{Kinemuchi12}.

As indicated in Fig.~\ref{ztf-lc} also during the
TESS observations IPHAS~J1908 remained in a quiescent
state. Prior to their analysis, the light curve was normalized by its 
median flux to allow the measurement of relative flux variations 
(such as eclipse depth) independently of the instrument's absolute calibration.
The resulting time series shows gaps associated with the TESS observing 
strategy (download of data), as well as more prominent 
discontinuities introduced by the PDCSAP processing. 
This reduces the effective observing time from 26.45~d to 16.67~d. The
noise level in the light curve of this faint (for TESS) source is too high to detect consistent variations. Therefore, we refain
from showing a graph of the light curve. 

\section{Timing investigation}
\label{Timing investigation}

In the following, we first report on our precise determination of the orbital period of IPHA~J1908 using both optical and X-ray data, then study the folded orbital light curves. We also report on our unsuccessful search for any spin period variation in the X-ray data.

\subsection{On the orbital period and ephemeries}

The frequent outbursts observed in the ZTF long-term light 
curve (Fig.~\ref{ztf-lc}) are typical for normal dwarf novae. The absence
of long lasting and brighter superoutbursts already indicates an orbital 
period above the CV period gap of $\sim$2--3~h, as is confirmed by the eclipse analysis 
(see below). Selecting only well sampled outbursts we estimate an average 
outburst amplitude of $\Delta zr = 2.3 \pm 0.2$ magnitude. This is quite common 
for dwarf novae with periods above 3~h. In a sample of 89 such systems,
\citet{Bruch24} found an average amplitude of $\Delta V = 2.5 \pm 1.0$ magnitude. 
We note that in the optical range the outburst amplitude does not depend 
strongly on the wavelength. The average outburst duration is $19 \pm 5$~d 
and the mean outburst interval is $53 \pm 15$~d. These values are also 
within the normal range for long period dwarf novae \citep{OtulakowskaHypka2016}. Thus, in its long-term behavior IPHAS~J1908 is 
not different from a typical dwarf nova.

The eclipsing nature of IPHAS~J1908 is apparent in the OPD observations. The light curve 
from the August 12, 2024 observations
(Table \ref{tab:opdobs}) covered two eclipses, whereas the others in 
the campaign from August 5 to 13 recorded one eclipse each. The ZE 
and PE observations on the night of August 5/6 were taken simultaneously 
and covered the same eclipse. To determine the orbital period, only the 
higher signal-to-noise light curve from the PE telescope was used.

In order to measure the orbital period, the eclipse epochs were determined 
by fitting a second-order polynomial to the eclipse profiles to find the 
minima. The signal-to-noise ratio of the light curves does not warrant 
the application of a more sophisticated method. Barycentric corrections 
were applied using the online tool provided by 
\citet{Eastman10}. The resulting eclipse epochs are listed in 
Table~\ref{tab:eclipse}, with the first eclipse observed at the PE assigned 
as cycle zero ($E=0$). In the absence of any cycle count ambiguities
they permitted to derive a preliminary period good enough to bridge the gap
to the previous observations with XMM-{\it Newton} and TESS.

\begin{table}[ht!]
\centering
\begin{tabular}{lccc}
\hline
Cycle & $T_{\rm min}$ & $(O-C)$ & Telescope \\
      & (BJD)       & (d)     &           \\
\hline
$-2265$           & 2460049.91587 &           $-0.00089$ & \phantom{T}XMM \\ 
$-2264$           & 2460050.12420 &           $-0.00393$ & \phantom{T}XMM \\
$-2263$           & 2460050.34411 & $\phantom{-}0.00459$ & \phantom{T}XMM \\
$-2262$           & 2460050.55244 & $\phantom{-}0.00150$ & \phantom{T}XMM \\
$-2261$           & 2460050.76077 &           $-0.00149$ & \phantom{T}XMM \\
$\phantom{0}-235$ & 2460479.01241 & $\phantom{-}0.00135$ &           TESS \\
$\phantom{-222}0$ & 2460528.68387 &           $-0.00066$ & \phantom{TX}PE \\
$\phantom{-222}4$ & 2460529.53102 & $\phantom{-}0.00098$ & \phantom{TX}ZE \\
$\phantom{-222}9$ & 2460530.58727 & $\phantom{-}0.00035$ & \phantom{TX}ZE \\
$\phantom{-22}28$ & 2460534.60235 &           $-0.00073$ & \phantom{TX}PE \\
$\phantom{-22}32$ & 2460535.44859 & $\phantom{-}0.00001$ & \phantom{TX}PE \\
$\phantom{-22}33$ & 2460535.66021 & $\phantom{-}0.00026$ & \phantom{TX}PE \\
$\phantom{-22}37$ & 2460536.50406 &           $-0.00140$ & \phantom{TX}PE \\
\hline
\end{tabular}
    \caption{Eclipse epochs}
    \label{tab:eclipse}
\end{table}

\begin{figure}[t!]
\figurenum{3}
\centering
\includegraphics[trim={0cm 0cm 0cm 0cm}, clip, width=\columnwidth]{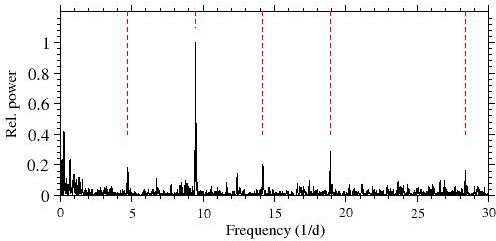}
   \caption{Lomb-Scargle periodogram of the TESS light curve of IPHAS~J1908.
The broken red lines indicate the orbital frequency and several of its harmonics.}
    \label{fig:tess}
\end{figure}

The TESS light curve is too noisy for individual eclipses to be 
identified. However, a Lomb-Scargle periodogram \citep{Lomb76,Scargle82}
 contains a peak at 4.731(5)~d$^{-1}$ (Fig.~\ref{fig:tess}). Well within the
formal error limits, determined using the method of 
\citet{1991MNRAS.253..198S}, the corresponding period is equal to the
preliminary orbital period of IPHAS~J1908. Other signals in the periodogram,
marked with red vertical bars, are all multiples of the fundamental orbital
signal within their respective error limits. Thus, there is no doubt that
they reflect harmonics of the orbital frequency. Folding the light curve on the
preliminary period now reveals the eclipses also in the TESS data. Choosing
the folding epoch such that the eclipse coincides with phase 0 yields a
representative eclipse epoch for the entire light curve. It is also listed
in Table~\ref{tab:eclipse} together with the corresponding cycle number.

Finally, timing analysis in X-rays was performed from light curves in three broad 
energy bands: 0.3--2\,keV, 2--10\,keV, and 0.3--10\,keV. Photon arrival 
times were converted to the Solar System barycenter and light curves were constructed from data binned into 400 s intervals. Because the X-ray data are too noisy 
for a formal fit to the eclipse profiles, the epochs were simply defined 
as the times of the minimum count rate. They are also included in 
Table~\ref{tab:eclipse}.

A linear least squares fit between the eclipse numbers and the 
corresponding epochs then yields the final ephemeris
\begin{equation}
T_{\rm min} = {\rm BJD}\, 2460528.6845(6) + 0.2113765(4) \times E
\end{equation}
where $E$ is the eclipse number. The numbers in parentheses denote the 
1$\sigma$ formal errors in the last decimal digits. The differences between 
the observed and calculated eclipse epochs, $(O-C)$, are listed in 
Table \ref{tab:eclipse}.

\begin{figure}[t!]
\figurenum{4}
\centering
\includegraphics[trim={1cm 0cm 3cm 6cm}, clip, width=\columnwidth]{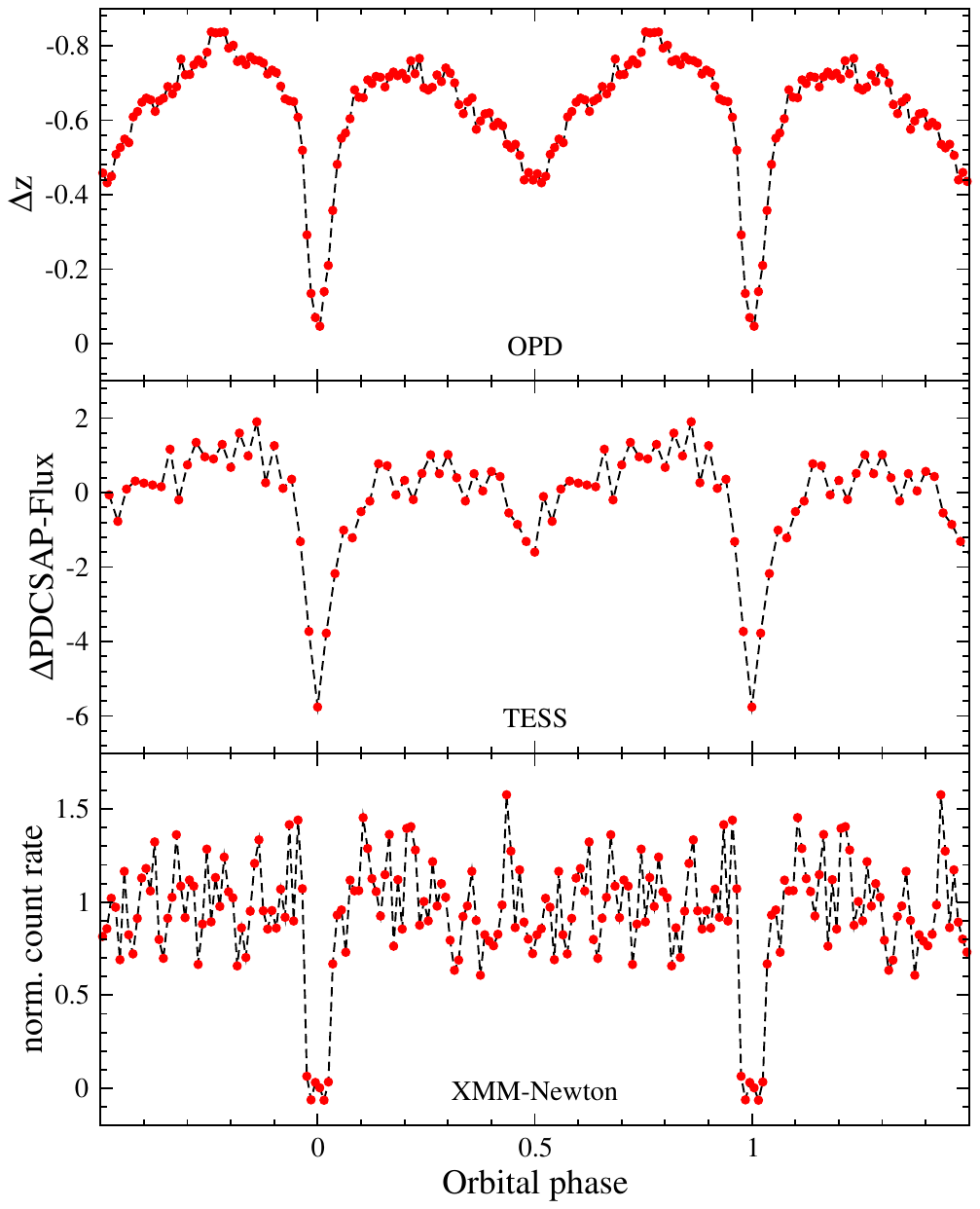}
   \caption{OPD (top), TESS (middle) and XMM-{\it Newton} (bottom) light curves
folded on the orbital period. The OPD and XMM-{\it Newton} data were binned in phase
intervals of 0.01, the TESS data in intervals of 0.02. In order to eliminate night-to-night
variations in the OPD light curves, the minimum magnitudes during eclipses 
was subtracted before averaging the nightly data.}
    \label{folded-lc}
\end{figure}

The phase folded OPD ($z$ band), TESS and XMM-{\it Newton} light curves are 
shown in Fig.~\ref{folded-lc}. The OPD data are binned in intervals of 0.01 in
phase. To eliminate the effects of night-to-night variations, the minimum 
magnitude during eclipses was subtracted from the individual light curves 
before averaging. Because of the lower S/N-ratio, a phase bin width of
0.02 was chosen for the TESS data. Not surprisingly, given the proximity
of the effective wavelength of the broad TESS pass-band and the OPD $z$ band, 
the orbital waveform is similar in both bands. It is dominated by a 
double-humped structure indicative of ellipsoidal variations, as expected 
for a system with such a long orbital period at long wavelengths. This also
explains the dominance of the first harmonic in the periodogram of the
TESS light curve. The slightly unequal maxima indicate that on top of the ellipsoidal variations and additional light source with a visibility depending on the orbital phase contributes to the total light, as is often seen in eclipsing long period CVs \citep[see, e.g.,][]{2024ApJS..273....6B}.

The folded X-ray light curve (using 0.01 cycle bins) of IPHAS~J1908 is reminiscent of that of HT~Cas \citep{Nucita09} in having a deep, sharp eclipse, and an approximately constant out-of-eclipse level. The X-ray eclipse of IPHAS~J1908 appears total.  The totality last for about 0.06 cycle, and the eclipse transition remains unresolved in our data ($<$0.01 cycle).  All these results are consistent with having a single, compact (with a size comparable to the white dwarf) X-ray emission region. While Z~Cha exhibits an energy-dependent out-of-eclipse variability and is interpreted as due to absorption by a bulge in the disk \citep{Nucita11}, any such effect is far less prominent in HT~Cas \citep{Nucita09}, so the observed characteristics of IPHAS~J1908 is within the range of behaviors exhibited by other deeply eclipsing dwarf novae.

\begin{figure}[ht!]
\figurenum{5}
\includegraphics[trim={1cm 8cm 3cm 10cm}, clip, width=\columnwidth]{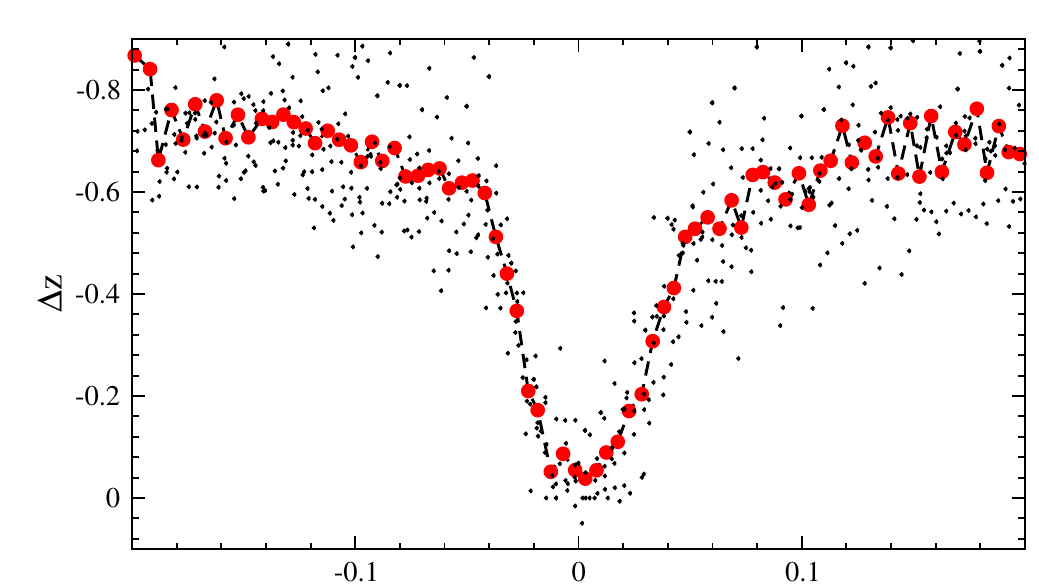}
   \caption{Detailed view of the optical eclipse profile. The red dots
            and black points represent average $r$ magnitudes (scaled to 
            a $\Delta r = 0$ at the eclipse minimum) and the original
            (unbinned) data points, respectively.}
    \label{ecl-profile}
\end{figure}

To obtain a reliable eclipse profile, the individual eclipses in the OPD
data were first carefully aligned in phase to eliminate small misalignments 
caused by the imperfect fits of the second-order polynomials originally used 
to define the epochs. For this purpose, the phase offset between individual 
profiles was determined as the maximum of their mutual cross-correlation 
functions. After subtracting this offset, the resulting average eclipse 
profile is shown in Fig. \ref{ecl-profile}. The black points represent 
the original data, while the red dots indicate average values binned in 
phase intervals of 0.005. The full width of the eclipse is only of the 
order of 0.09 in phase. This is significantly less than observed in most
deeply eclipsing CVs and suggests a small accretion disk.

\subsection{Search for fast coherent pulsations}

In order to look for the presence of coherent variations hidden in the
data noise, Lomb-Scargle periodogram was applied to all light curves, 
sampling all frequencies up to the Nyquist frequency. No such variations other than orbital were detected in the optical light curves. 
To investigate fast pulsations, even in the low signal-to-noise regime, 
we explore the X-ray data. XMM-{\it Newton} EPIC/pn
event lists were analyzed using the $Z^{2}_{n}$ periodogram \citep{Buccheri83}.  To achieve this, events were extracted from a circular region with radius of only 12.5 arcsec centered on the source position. The search was done from the Nyquist limit (2/$T_{exp}$) of 2.148$\times$10$^{-5}$ Hz ($\sim$\,46,550\,s) up to 0.05\,Hz (20\,s), oversampling by 10$^3$ in steps of $\sim$\,1.1$\times$10$^{-8}$ Hz. Because the low signal-to-noise, we discuss here only results from the pn camera. 

No peak in the $Z^{2}_{n}$ periodogram corresponds to the orbital period of the system. However, the three prominent peaks are at 8,883\,s, 3,047\,s, and 1,521\,s, which correspond to 1/2, 1/6, and 1/12 of the orbital period, respectively. These peaks are formally associated with pulsed fraction (PF) of $\sim$\,19\%, 18\%, and 16\%, respectively. The prominence of these orbital harmonics is consistent with the shape of the orbitally-folded X-ray light curve, containing a sharp eclipse but otherwise flat. Moving to higher frequencies, no significant peaks are detected. To estimated an upper limit on the PF of any potential modulation in the data, we adopt the most significant peak, located at a period of 129.7\,s, whose power corresponds to a PF of $\sim$\,14\%. 

\section{Spectroscopic investigation}

\subsection{Optical spectroscopy}

The GMOS/Gemini South optical spectroscopy 
(Fig.~\ref{fig:opticalspectrum}) reveals a continuum characterized 
by intense hydrogen Balmer emission lines and 
numerous lines of He\,\textsc{I} and Fe\,\textsc{II} transitions. 
In general, lines exhibit distinct double-peaked profiles. This is the
classical signature of a (Keplerian) accretion disk. The high-excitation 
He\,\textsc{II}\,$\lambda$4686 line is also
detected, but is weak. Table~\ref{tab:optlines} lists the equivalent
width (EW) and the full width at half maximum (FWHM) of the main emission lines. 
The line parameters were determined by fitting Gaussian profiles to the normalized continuum. The reported 1$\sigma$ uncertainties correspond to the formal errors derived from the covariance matrix of the fit.

\begin{table}[ht]
    \centering
    \hspace*{-0.5in}
    \begin{tabular}{ccc}
         \hline
         \hline
         Line & EW  & FWHM\\
         & (\AA) &  (\AA)\\
         \hline
         HeII 4686 & 5.98 $\pm$ 1.48 & 29.97 $\pm$ 5.59\\ [2pt]
         H$\alpha$ & 46.35 $\pm$ 2.56 & 22.29 $\pm$ 1.07\\ [2pt]
         H$\beta$  & 34.89 $\pm$ 2.16 & 23.94 $\pm$ 1.12\\ [2pt]
         H$\gamma$ & 30.16 $\pm$ 1.52 & 24.55 $\pm$ 0.94\\ [2pt]
         H$\delta$ & 16.45 $\pm$ 1.59 & 18.05 $\pm$ 1.91\\ [2pt]
         H$\epsilon$ + Ca\,\textsc{II} 3968 & 20.19 $\pm$ 3.72 & 16.77 $\pm$ 2.87 \\ [2pt]
         H$\zeta$ &  17.02 $\pm$ 2.32 & 22.21 $\pm$ 1.79\\ [2pt]
         He I 6678 & 3.37 $\pm$ 0.85 & 17.81 $\pm$ 2.74\\ [2pt]
         He I 5876 & 4.59 $\pm$ 2.14 & 15.79 $\pm$ 5.72\\ [2pt]
         He I 4921 + Fe\,\textsc{II} 4924 & 2.08 $\pm$ 1.03 & 15.32 $\pm$ 5.74\\ [2pt]
         He I 4471 & 2.62 $\pm$ 0.62 & 10.66 $\pm$ 1.89\\ [2pt]
         Fe II 5169 & 2.09 $\pm$ 0.57 & 17.19 $\pm$ 3.35\\ [2pt]
         \hline
    \end{tabular}
    \caption{Measured EW and FWHM values for optical spectral lines.}
    \label{tab:optlines}
\end{table}

\subsection{X-ray spectroscopy}

\begin{figure}[t!]
\figurenum{6}
\includegraphics[trim={2cm 1cm 2.5cm 2cm}, clip, width=\columnwidth]{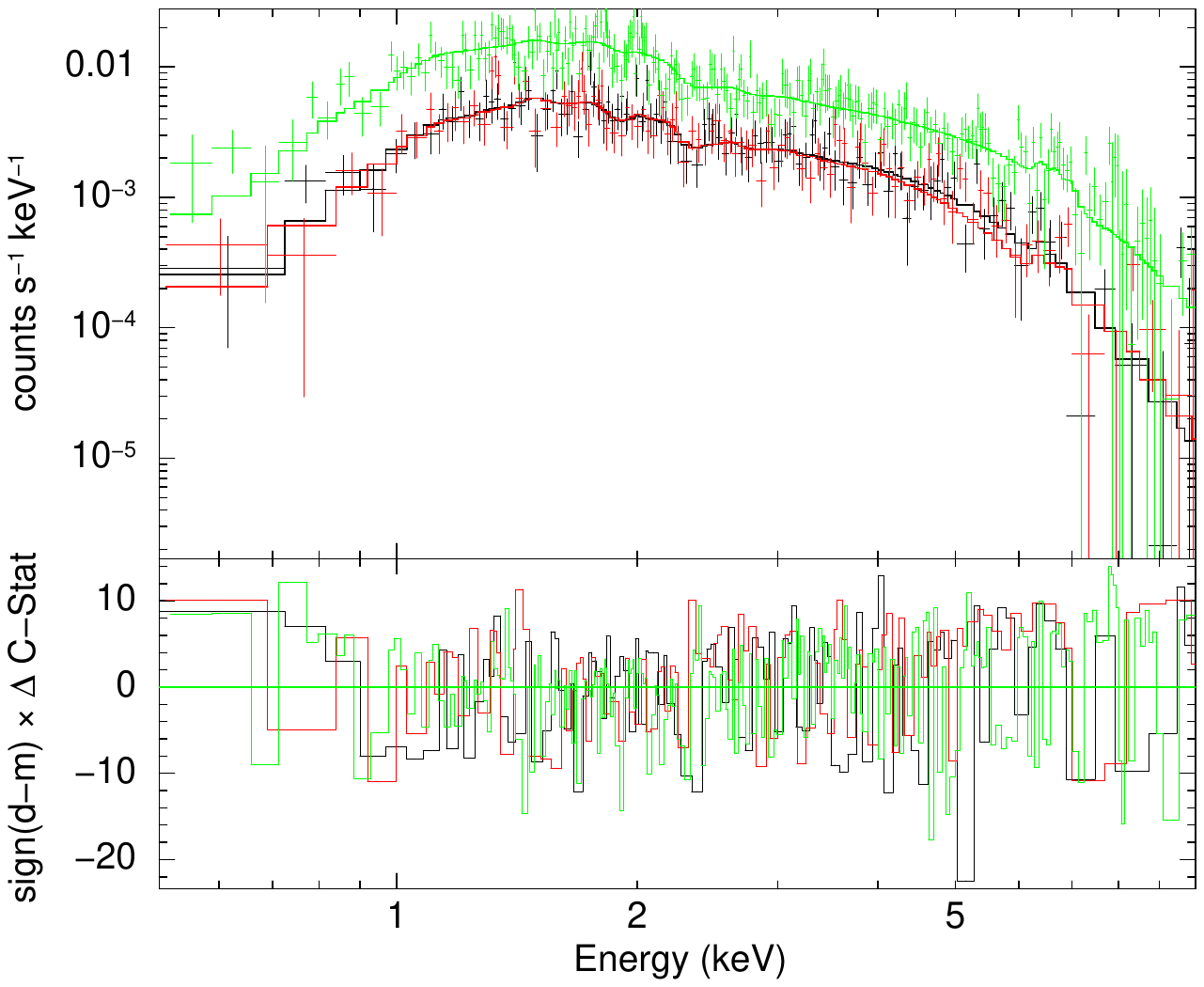}
   \caption{XMM-{\it Newton} X-ray spectra (top panel) and residuals (bottom). Green, black, and red colours correspond to the pn, MOS1, and MOS2 data (crosses), and their best simultaneous spectral fit and residuals (solid lines), respectively.}
    \label{fig:xrayspectra}
\end{figure}

To characterize the high-energy emission of IPHAS~J1908, we analyzed the 
time-averaged XMM-{\it Newton} spectra (Fig. \ref{fig:xrayspectra}). 
To this end, source and background spectra, along with 
their respective response matrices 
and ancillary response files, 
were generated. The energy channels were grouped to a minimum of 25 counts 
per bin to allow the use of $\chi^2$ statistics. Spectral fitting was 
performed simultaneously on the three EPIC spectra using XSPEC v12.15.1. 
We conservatively restricted the fitting to the 0.6--7.8\,keV energy 
band because the background contribution degrades the spectrum beyond these limits.
And following the complexity of accreting WDs, the applied models evolved from a basic single-temperature optically thin thermal plasma (\textsc{apec}) with simple photoelectric absorption (\textsc{tbabs}), to incorporating complex partial-covering absorption (\textsc{partcov$\times$tbabs}, with the parameter \texttt{CvrFract} of \textsc{partcov} varying from 0 to none, up to 1 for a total coverage), and ultimately progressing to a cooling flow model (\textsc{mkcflow}), with a gaussian line (\textsc{gauss}) to account for the fluorescent iron line at 6.4\,keV. Thus, the best simplest description to describe the X-ray spectrum was \textsc{constant} $\times$ \textsc{tbabs} $\times$ (\textsc{partcov} 
$\times$ \textsc{tbabs}) $\times$ (\textsc{apec/mkcflow} + \textsc{gauss}). 
The best-fitting parameters with the \texttt{apec} and \textsc{mkcflow} models are summarized in Table~\ref{tab:xrays}.

The data description with a single-plasma model yielded a reduced chi-squared of 
$\chi^2_{\nu} = 1.09$ for 224 degrees of freedom (see Table \ref{tab:xrays}). The fully covering 
interstellar column density is constrained to 
$N_{H} = 0.90^{+0.08}_{-0.09} \times 10^{22}$\,cm$^{-2}$. In addition, we detect 
a significant intrinsic partial covering absorber affecting a fraction of 
$0.40 \pm 0.07$ of the X-ray emitting region, with a high localized 
column density of $N_{H} = 5.6^{+3.4}_{-2.1} \times 10^{22}$\,cm$^{-2}$. The primary continuum is well modeled by a single-temperature thermal 
plasma with $kT = 7.8^{+1.4}_{-0.9}$\,keV. To 
account for the Fe\,K$\alpha$ fluorescence line, we included a narrow Gaussian component 
($\sigma$ fixed at $1 \times 10^{-6}$\,keV) at a fixed rest energy of 
$6.4$\,keV. Although the line flux is modestly constrained to 
$F_{\rm line} = (4.5 \pm 7.1) \times 10^{-7}$\,photons\,cm$^{-2}$\,s$^{-1}$, 
its inclusion is physically motivated -- as a feature ubiquitously 
produced by the irradiation of the WD surface and, in the case of IPs, from the pre-shock flow by the hot column. Finally, the fit yields an unabsorbed X-ray flux of 
$F_{X\,;\,0.3-10\,keV)}$ $\sim$ 1.3 $\times 10^{-12}$\,erg\,cm$^{-2}$\,s$^{-1}$. Adopting 
a geometric distance of $d = 1076^{+181}_{-124}$\,pc \citep[derived from 
\textit{Gaia} astrometry;][]{Bailer-Jones21}, this corresponds to an intrinsic X-ray 
luminosity of $L_{X\,;\,0.3-10\,keV} \sim 1.7 \times 10^{32}$\,erg\,s$^{-1}$. 

Because the effective area of the XMM-{\it Newton} EPIC cameras drops 
significantly above $\sim$\,8\,keV, fits to multi-temperature cooling flow 
models often stall near the edge of the instrumental bandpass, failing 
to describe the hottest emitting regions of the shock. In fact, the maximum temperature of the \textsc{mkcflow} model $kT_{max}$ explored in this work is unconstrained: the value converges to $kT_{max} \sim 26$\,keV during the fit, but becomes unconstrained when calculating the uncertainties with the \texttt{ERROR} command, yielding only a lower limit of $kT_{max} > 17$\,keV. The results replacing the \texttt{apec} by the \texttt{mkcflow} component, freezing $kT_{max}$ in 26\,keV (and $kT_{min}$ in 0.0808\,keV), are shown in Table \ref{tab:xrays}. Consequently, the true shock temperature may be significantly higher. Nevertheless, the best-fit $kT_{max}$ value of $\sim 26$\,keV for IPHAS~J1908 appears reasonable: when the $<$10 keV X-ray spectra of a quiescent dwarf nova is fit with a single-temperature plasma model with $kT \sim 8$ \,keV, the best-fit $kT_{max}$ value for a \textsc{mkcflow} model fit is 20--25\,keV (compare Tables 4 and 5 of \citealt{Byckling10}). Under such an assumption, the accretion rate derived from the \texttt{mkcflow} model is consistent with 
$\dot{M} \sim$ 4.5$\times$10$^{-11}$ M$_{\odot}$\,yr$^{-1}$. 
As for IPs, not only is a cooling flow expected to explain the X-ray emission, but the inclusion of a reflection component also proves necessary \citep[e.g.,][]{2019ApJ...880..128L}. 

\begin{table*}[ht!]
\centering
    \begin{tabular}{cccc}
         \hline
         \hline
         \multicolumn{4}{c}{ \textsc{constant$*$tbabs$*$(partcov$*$tbabs)$*$(apec/mkcflow+gauss)}}\\
         \hline
         Component & Parameter & \multicolumn{2}{c}{Value}\\
         \hline
         \textsc{tbabs} & N$_{H}$ (10$^{22}$\,cm$^{-2}$)    & 0.90$^{+0.08}_{-0.09}$ & 1.11$^{+0.09}_{-0.13}$ \\
         \hline
         \textsc{partcov} &   CvrFract  & 0.40$^{+0.06}_{-0.07}$ & 0.35$^{+0.09}_{-0.05}$\\
         \textsc{tbabs} & N$_{H}$ (10$^{22}$\,cm$^{-2}$)    & 5.6$^{+3.4}_{-2.1}$ & 6.3$^{+5.8}_{-3.0}$\\
         \hline
         \textsc{apec}  & $kT$ (keV)                        & 7.8$^{+1.4}_{-0.9}$ & ... \\
                        & normalization                      & (6.5$^{+0.6}_{-0.5}$)$\times$10$^{-4}$ & ...\\
        \hline  
        \textsc{mkcflow} & $kT_{max}$                        & ... & 26$^*$ \\
          & normalization (M$_{\odot}$\,yr$^{-1}$)                                  & ... & (4.5$\pm$0.3)$\times$10$^{11}$ \\                 
        \hline
         \textsc{gauss} & E$_{c}$ (keV)                     & 6.4$^*$ & 6.4$^*$\\
                        & $\sigma$ (keV)                    & 1$\times$10$^{-6}$$^*$ & 1$\times$10$^{-6}$$^*$ \\
                        & flux (photons\,cm$^{-2}$\,s$^{-1}$)                 & $<$1.1$\times$10$^{-6}$ & $<$1.0$\times$10$^{-6}$ \\
         \hline
         & unabs. 0.3-10 keV flux (erg\,cm$^{-2}$\,s$^{-1}$) & (1.2-1.3)$\times$10$^{-12}$ & (1.3)$\times$10$^{-12}$ \\
         \hline
         & $\chi$$^2_{\nu}$/d.o.f. & 1.09/224 & 1.10/225\\ 
         \hline
    \end{tabular}
    \caption{Best-fit parameters for the XMM-{\it Newton} X-ray spectral analysis.}
    \label{tab:xrays}
\end{table*}

\section{Discussion}
\label{Discussion}

\subsection{On the secondary star}
\label{subsct:opticalproperties}

Apart from the orbital period determination, the optical eclipses observed in IPHAS~J1908 provide two immediate pieces of information (see Fig.~\ref{ecl-profile}). 
The first is that the gradual decrease in brightness at ingress and the gradual increase in brightness at egress indicate the presence of an accretion disk -- corroborated from the presence of double-peaked profiles in emission lines (see Fig. \ref{fig:opticalspectrum}). 
The second is that the bottom of the eclipse is 
approximately flat, indicating that during the eclipse minimum, the primary 
component is completely obscured. Thus, the magnitude at minimum reflects 
the brightness of the secondary star, serving as a reliable upper limit 
that is likely close to its true value. This provides a lever to 
estimate several parameters of the secondary star. 

IPHAS~J1908 suffers from considerable interstellar extinction. Using
3D extinction maps \citep{Doroshenko24,Edenhofer24}\footnote{
\url{http://astro.uni-tuebingen.de/nh3d/nhtool}}, we estimate a visual 
extinction of $A_V = 2.1 \pm 0.6$\,mag at the distance of the system. 
From the wavelength-dependent extinction law of 
\citet{Cardelli89},
the absorption at the effective wavelength of the Sloan $r$ band 
(6231\,\AA)\footnote{
\url{https://skyserver.sdss.org/dr1/en/proj/advanced/color/sdssfilters.asp}} 
is $A_r = 1.86 \pm 0.53$. The difference between the differential 
magnitude of IPHAS~J1908 and the average magnitude of the ensemble of 
comparison stars used in the PE photometry is $\Delta r = 16.40$. 
This allows us 
to convert the PE light curves from differential to calibrated magnitudes. 
The magnitude of IPHAS~J1908 during the eclipse minimum then is
on average of $r = 20.61 \pm 0.11$. 
The absolute $r$-band magnitude of the secondary star is thus estimated 
to be $M_{r,{\rm s}} = 8.59 \pm 0.62$, where the error is propagated from the 
uncertainties in the apparent magnitude, distance, and interstellar 
absorption (with the latter dominating the error budget). Note that,
strictly speaking, this value should be regarded as an upper limit to the 
brightness of the secondary star.

To assess the validity of this result, we compared it to the semi-empirical 
secondary star sequence of \citet{Knigge11}. Given the orbital period of 
5.073\,h for IPHAS~J1908, its secondary is expected to have a 
Johnson-Cousins $R$-band magnitude of 8.56 \citep[see Table 2 of][]{Knigge11}. 
Accounting for the difference in effective wavelengths between the $r$ and 
$R$ bands, and interpolating between $V$ and $R$ in their table, the 
expected $r$-band magnitude is 8.84. This is comfortably within the 1$\sigma$ 
uncertainty of our calculated absolute magnitude. Moreover, it is fully 
consistent with $M_{r,{\rm s}}$ being an upper limit to the brightness. Based on the \citet{Knigge11} sequence at the orbital period of IPHAS~J1908, 
we can also infer other expected parameters for the secondary star,
namely, a mass of 0.50\,$M_\odot$, a radius of 0.55\,$R_\odot$, an 
effective surface temperature of 3700\,K, and a spectral type of M2.3.

Out-of-eclipse, the average magnitude of IPHAS~J1908 is 1.04~mag
brighter than during eclipse minimum in the $r$ band. Thus, the absolute 
magnitude is $r = 7.63$. Neglecting the (small) typical difference of 
$V-r \sim 0.1\, {\rm to}\, 0.2$ for long period dwarf novae, this is about 
a magnitude fainter than the maximum of the broad absolute magnitude 
distribution of dwarf novae above the CV period gap \citep{Bruch24}, 
consistent with the dwarf nova-type outburst behavior in the long time
scale light curve.

The high intrinsic luminosity of IPHAS~J1908 strongly indicates the presence 
of a bright, dominant accretion disk, which is inherently consistent with 
the relatively long orbital period of 5.073\,h determined from our OPD 
photometry. The presence of a high-inclination accretion disk is further 
supported by the relatively broad, double-peaked emission lines observed 
in the optical spectrum (Fig. \ref{fig:opticalspectrum}). Finally, 
the high intrinsic luminosity inferred from the absolute magnitude is a third argument supporting the presence of a disk.

\subsection{Learning from the X-ray emission}

Even though both the light curve (Fig. \ref{ztf-lc}) and the optical spectrum (Fig. \ref{fig:opticalspectrum}) of IPHAS~J1908 are consistent with a dwarf nova classification, its X-ray properties make it an outlier for dwarf novae. First, IPHAS~J1908 is more X-ray luminous than the sample of dwarf novae studied by \cite{Byckling10}.\footnote{Note that the X-ray luminosity values for SS~Cyg tabulated in Table 6 of \cite{Byckling10} are based on the then current distance estimate of 165 pc. With the \citet{Bailer-Jones21} distance of 112.3 pc, the 2--10 keV luminosity of SS~Cyg is 6.9$\times 10^{31}$ erg\,s$^{-1}$ meaning no dwarf novae in their sample have an 2--10 keV X-ray luminosity above 10$^{32}$ erg\,s$^{-1}$ (see their Figure 4). Extrapolated to 0.3--10 keV, the X-ray luminosity of SS~Cyg is about 1.1$\times 10^{32}$ erg\,s$^{-1}$, lower than that of IPHAS~J1908.} In addition, its X-ray spectrum is unusually hard for a dwarf nova, and requires a high mass WD.  The inferred $kT_{max}$ of $\sim 26$ \,keV requires a high-mass WD, perhaps above 1.0 M$_\odot$ (see \citealt{Yu18,Mukai22}). These X-ray properties may make IPHAS~J1908 an extreme dwarf novae, worthy of further studies.  Before proceeding further with the interpretation, it is therefore worth investigating if it might belong to a different CV subclass.

Some non-magnetic CVs usually remain in an optically bright state similar to dwarf novae in outbursts: these are the nova-like systems. Their X-ray properties are poorly understood, but some may reach or exceed luminosities of 10$^{32}$ erg\,s$^{-1}$ (Islam \& Mukai, in preparation). However, the ground-based photometry of IPHA~J1908 indicates it not to be a nova-like system.  We therefore consider the possibility of IPHAS~J1908 being a magnetic CV, which often have higher X-ray luminosities.

Since optical spectroscopy and photometry clearly rule out a polar nature for this system (see Section \ref{subsct:opticalproperties}), we will concentrate on the IP scenario. The X-rays in IPs are produced in one (or two) accretion column(s), or rather longitudinally extended curtains, and the absorption naturally arises from the dense, pre-shock material in the accretion curtains 
periodically crossing the line of sight. 
Most well-known IPs are both optically bright (similar to nova-like systems) and X-ray luminous \citep[$log L_x (erg\,s^{-1}) > 32.8$;][]{Mukai23} -- hereafter high luminosity IPs (HLIPs). Hard ($kT_{max} > 20$ keV) X-ray spectrum and the presence of complex absorber are hallmarks of these normal IPs (see, e.g., \citealt{2019ApJ...880..128L}). While the X-ray spectrum of IPHAS~J1908 is hard and requires a partial-covering absorber for a good fit, similar to these standard IPs, its X-ray luminosity ($1.7 \times 10^{32}$ erg\,s$^{-1}$) is lower than those of HLIPs.  Instead, IPHAS~J1908 may be a low-luminosity IP (LLIP), a separate subclass of systems with an optically faint disk and a lower X-ray luminosity  ($\lesssim$\,10$^{32}$\,erg\,s$^{-1}$).\footnote{https://asd.gsfc.nasa.gov/Koji.Mukai/iphome/catalog/llip.html} In HLIPs, the specific accretion rate (the accretion rate per unit area) is high and hence the accretion column is shocked close to the WD surface. In this case, the maximum temperature ($kT_{max}$) is a good measure of the WD mass, expected to be 0.65 M$_\odot$ for $kT_{max}$ of 26 keV. On the other hand, the shock in LLIPs may form well above the WD surface; in this interpretation, the inferred $kT_{max}$ of IPHAS~J1908 may well be consistent with a $\sim$0.8 $M_\odot$ WD, depending on the actual shock height (cf. the case of EX~Hya; \citealt{Luna15}). 

In our view, the X-ray spectrum of IPHAS~J1908 is consistent with the LLIP interpretation or the dwarf nova interpretation. 
The X-rays from dwarf novae come from the boundary layer, at the limit between the inner part of the accretion disk and the WD surface, with the complex absorption being produced by the material involving the system near to an edge-on configuration. 
In the dwarf nova scenario, the X-ray observation of IPHAS~J1908 caught it in a quiescent state.  Note that, while a detectable level of partial-covering absorber is rare, it is known in high-inclination dwarf novae \citep[e.g., V893 Sco;][]{2009ApJ...707..652M}. 
Also, the X-ray luminosity is consistent with the high end of the distribution for dwarf novae.

\subsection{About the nature of the system}

 The presence of an accretion disk, the weakness of the He\,\textsc{ii}\,$\lambda$4686 line, and the frequent dwarf
nova-like outbursts, exclude the possibility of the system being a polar (AM Herculis). The frequency of optical outbursts also discards ordinary nova-like systems. However, while the optical features alone firmly argue against a strongly magnetic polar
scenario and nova-like, they favors the dwarf nova scenario but an IP nature cannot be conclusively ruled out.

The weakness of He\,\textsc{ii}\,$\lambda$4686 line, while not strictly conclusive on its own, suggests a 
weakly magnetic WD and is consistent with both IP and dwarf novae scenario \citep{Patterson1994,Bloemen2010,Szkody2002}. On the other hand, the absence of indications for WD spin variations in the optical and X-ray power spectra argues against the IP scenario but is not conclusive. Such variations could be intrinsically invisible by geometrical factors, the period might lie above the Nyquist frequency, or the detection could be compromised by either a short time baseline or a low signal-to-noise ratio. This limit is below 4~min for the TESS data and 2~min for the other optical data; however, the latter's time baseline is significantly shorter, making detection more difficult. In X-rays, any period determination is compromised by the low signal-to-noise ratio, and the pulsed fraction is upper-limited to less than 14\%.  

Finally, the ZTF light curve (Fig. \ref{ztf-lc}) provides a 
$\sim$\,7-year baseline to examine the system's long-term variability. For one thing, it provides a statistical view of the system's outburst behavior. The outburst frequency of IPHAS~1908 is significantly higher than that of other IPs, which typically exhibit not only less frequent, but also even shorter-lived outbursts.
On the other hand, their frequency and duration in the system are quite normal for most dwarf nova.
This behavior can be physically interpreted through two plausible 
scenarios common in IPs. 
First, the truncation of the 
inner accretion disk by the WD's magnetic field alters the 
classical disk instability model \citep[named DIM; see][]{2001NewAR..45..449L}, often 
resulting in short-lived, low-amplitude outbursts that are easily missed 
by low-cadence surveys \citep{Hameury86, Hellier00}. 

The other ZTF result is that the quiescence level of the system fluctuates at a level of a few tenths of a magnitude over long time scales. Such fluctuations have been seen in dwarf novae -- e.g., the 2021 anomalous state of SS Cyg reported by  \citet{Kimura21}, which these authors interpreted as due to changes in viscosity of the disk. Alternatively, the quiescent level fluctuation in IPHAS~J1908 may be related to the low states occasionally seen in IPs \citep{Covington22} and some novalike CVs \citep{2004AJ....128.1279H}.  Although uncertain, changes in mass transfer rate from the donor is often invoked as a possible explanation \citep{Garnavich88, Hessman00, Kafka04}. 

\section{Final remarks}

IPHAS~J190812.63+045728.1 stands out from the general population of 
cataclysmic variables due to a rare combination of observational features. 
The evidence points to either a DN nature, as favored by the outburst frequency and duration, and that rule out the novalike classification, or an LLIP, as supported by modestly luminous, hard-thermal X-rays. Thus, rather than being a 
standard addition, the system serves as a unique precision 
laboratory that challenges and refines our understanding of accretion disks. The novelty on IPHAS~J1908 is primarily anchored in three key aspects:

\begin{itemize}

\item First, IPHAS~J1908 exhibits rare, flat-bottomed total eclipses in 
X-rays and in the optical. In typical CVs, the persistent 
glare of the accretion disk heavily contaminates the photometric and spectral 
signatures of the donor star. In this system, however, the secondary 
completely occults both the primary WD and the entire emitting region of the disk. This natural geometry provides a pristine, 
model-independent window to characterize the donor star, allowing us to 
constrain it as an M2.3 dwarf with a mass of $\sim 0.50\,M_\odot$. Such an 
uncontaminated view is rare and makes this system a benchmark 
for testing CV evolutionary models at relatively long orbital periods 
($P_{\rm orb} \sim 5.073$\,h).

\item Second, regardless of whether it is a dwarf novae or an IP, it is an unusual system for its combination of its X-ray luminosity and the frequent outbursts with typical dwarf nova characteristics. If IPHAS~J1908 is a dwarf nova, it sits at the extreme high end of X-ray luminosity range for this class.  Follow-up observations, including in X-rays, are warranted to investigate the physical reason behind the high X-ray luminosity. If, on the other hand, IPHAS~J1908 is an IP, the reason why this IP can exhibit apparently typical dwarf nova outbursts, suggestive of thermal-viscous disk instability, needs to be investigated. The presence of these outbursts in a relatively long-period IP, combined with the geometric constraints provided by its total eclipses, offers the opportunity to study how a magnetic field truncates the inner disk and affects the origin and the propagation of heating waves during an outburst cycle. In addition, if it is an IP, the weakness of its He\,\textsc{ii}\,$\lambda$4686 line is an interesting topic of research, and also have a practical consequence in that the canonical optical selection criteria for magnetic CVs are incomplete, as they can miss some IPs.

\item Finally, the long-term archival photometry from ZTF reveals that 
the system undergoes fluctuations on the order of 
a few tenths of a magnitude during quiescence. While such fluctuations are known both in dwarf novae and in IPs, there is yet no secure and detailed physical explanation.  This is another area where more data are essential for a better understanding.
\end{itemize}

Together, these properties render IPHAS~J1908 a prime target for future 
follow-up, including high-cadence eclipse mapping, hard X-ray spectroscopy 
($>20$\,keV), and time-resolved ultraviolet observations.

\begin{acknowledgments}
R.L.O. was partially supported by {\it Conselho Nacional de Desenvolvimento Científico e Tecnológico} (CNPq PQ-315632/2023-2 and 445047/2024-0; Brazil) and acknowledges financial support from NASA and the XRISM Project Science Office during his stay at NASA’s Goddard Space Flight Center. A.S.A. was supported by the {\it Coordenação de Aperfeiçoamento de Pessoal de Nível Superior - Brasil} (CAPES) - Finance Code 001 (88887.992353/2024-00). 
G.P.G. was supported by CNPq (157556/2024-7). 
Partially based on observations obtained at the Pico dos Dias Observatory (OPD/LNA), operated by the Laboratório Nacional de Astrofísica (Brazil), under observing programs 2024A-I030 (PI: L.\ Almeida) and 2024-I-039 (PI: A.\ Bruch). 
Partially based on observations obtained at the international Gemini Observatory (DDT program 
GS-2024B-DD-103; PI: R. Lopes de Oliveira), a program of NSF NOIRLab, which is managed by the Association of Universities for Research in Astronomy (AURA) under a cooperative agreement with the National Science Foundation on behalf of the Gemini partnership: the National Science Foundation (United States), National Research Council (Canada), Agencia Nacional de Investigación y Desarrollo (Chile), Ministério da Ciência, Tecnologia e Inovação (Brazil), Ministerio de Ciencia, Tecnología e Innovación (Argentina), and Korea Astronomy and Space Science Institute (Republic of Korea). 
This research paper includes data collected by the TESS mission, which are publicly available from the Mikulski Archive for Space Telescopes (MAST) at the Space Telescope Science Institute (STScI). Funding for the TESS mission is provided by NASA's Science Mission Directorate. STScI is operated by the Association of Universities for Research in Astronomy, Inc., under NASA contract NAS 5-26555. 
Partially based on observations obtained with the Samuel Oschin 48-inch Telescope at the Palomar
Observatory as part of the Zwicky Transient Facility project. ZTF is supported by the National
Science Foundation under Grant No. AST-1440341 and a collaboration including Caltech, IPAC,
the Weizmann Institute for Science, the Oskar Klein Center at Stockholm University, the University
of Maryland, the University of Washington, Deutsches Elektronen-Synchrotron and Humboldt
University, Los Alamos National Laboratories, the TANGO Consortium of Taiwan, the University
of Wisconsin at Milwaukee, and Lawrence Berkeley National Laboratories. Operations are
conducted by COO, IPAC, and UW.
\end{acknowledgments}

\vspace{5mm}
\facilities{S-PLUS, Gemini (GMOS), OPD/LNA, TESS, ZTF.}

\software{astropy \citep{2013A&A...558A..33A,2018AJ....156..123A},  
          Xspec \citep{1996ASPC..101...17A},
          Dragons \citep{2023PASP..135g4502L},
          MIRA \citep{Bruch1993}
          }

\bibliography{sample631}{}

\end{document}